\documentclass[a4paper,12pt]{article}
\pdfoutput=1
\usepackage{jheppub}
\usepackage{amsmath,amssymb,bm}
\usepackage{graphicx}
\usepackage{hepunits}
\usepackage[svgnames]{xcolor}
\usepackage{multirow}
\usepackage{float}
\usepackage{textcomp}

\allowdisplaybreaks

\title{Mixed QCD-EW corrections for Higgs leptonic decay via $HW^+W^-$ vertex}
\author[a]{Chichuan Ma,}
\author[a]{Yuxuan Wang,}
\author[b]{Xiaofeng Xu,}
\author[c]{Li Lin Yang,}
\author[a]{Bin Zhou}

\affiliation[a]{School of Physics and State Key Laboratory of Nuclear Physics and Technology,\\
Peking University, Beijing 100871, China}
\affiliation[b]{Institut f\"ur Theoretische Physik, Universit\"at Bern, Sidlerstrasse 5, CH-3012 Bern, Switzerland}
\affiliation[c]{Zhejiang Institute of Modern Physics, Department of Physics, Zhejiang University, Hangzhou 310027, China}

\emailAdd{chichuanma@pku.edu.cn}
\emailAdd{wangyuxuan119@pku.edu.cn}
\emailAdd{pkuxxf@gmail.com}
\emailAdd{yanglilin@zju.edu.cn}
\emailAdd{2028910235@pku.edu.cn}

\abstract{We consider the two-loop corrections to the $HW^+W^-$ vertex at order $\alpha\alpha_s$. We construct a canonical basis for the two-loop integrals using the Baikov representation and the intersection theory. By solving the $\epsilon$-form differential equations, we obtain fully analytic expressions for the master integrals in terms of multiple polylogarithms, which allow fast and accurate numeric evaluation for arbitrary configurations of external momenta. We apply our analytic results to the decay process $H \to \nu_e e W$, and study both the integrated and differential decay rates. Our results can also be applied to the Higgs production process via $W$ boson fusion.}

\begin{document}

\maketitle

\clearpage

\section{Introduction}

After the discovery of the Higgs boson in 2012, one of the primary goals of current and future high energy collider experiments is the precision measurements of various Higgs properties. These include the Large Hadron Collider (LHC) and its future upgrade High-Luminosity LHC (HL-LHC) \cite{Apollinari:2017cqg}, as well as future electron-positron colliders such as the Circular Electron Position Collider (CEPC) \cite{CEPCStudyGroup:2018ghi}, the Future Circular Collider (FCC-ee) \cite{Abada:2019lih} and the International Linear Collider (ILC) \cite{Bambade:2019fyw}. An important property of the Higgs boson is its coupling to weak gauge bosons, namely, the $Z^0$ boson and the $W^\pm$ boson. Their precision knowledge is crucial for verifying the spontaneous electroweak symmetry breaking mechanism and for exploring the nature of the Higgs sector. If the Higgs sector is more complicated than the simplest version assumed by the Standard Model (SM), e.g., if the Higgs boson is composite, these gauge couplings may differ from the SM value by a tiny amount. The precision measurement of these couplings will then help us to detect such small effects resulting from physics at higher scales.

With the current LHC data, the $HZZ$ coupling and the $HW^+W^-$ coupling have been measured with an uncertainty of 11\% and 15\%, respectively \cite{Sirunyan:2018koj}. At HL-LHC, the precision is estimated to be about 1.5\% and 1.7\%, respectively \cite{Cepeda:2019klc}. At future electron-position colliders, the precisions for these couplings can be greatly improved. For the $HZZ$ coupling, the accuracy is estimated to be about 0.25\% at the CEPC with a center-of-mass energy $\sqrt{s} = \unit{250}{\GeV}$ \cite{CEPCStudyGroup:2018ghi}, and about 0.17\% at the FCC-ee with $\sqrt{s} = \unit{365}{\GeV}$ \cite{Abada:2019lih}; while for the $HW^+W^-$ coupling, the precision is estimated to be about 1.4\% at the CEPC and about 0.43\% at the FCC-ee. To get the most out of the future high precision data, it is necessary to provide accurate theoretical predictions for production and decay processes sensitive to these couplings. Those related to the $HZZ$ vertex have been extensively studied in the literature \cite{Fleischer:1980ub, Fleischer:1982af, Kniehl:1991hk, Denner:1992bc, Kniehl:1995ra, Frink:1996sv, Djouadi:1997rj, Kniehl:2012rz, Gong:2016jys, Sun:2016bel, Chen:2018xau, Wang:2019fxh}. In this paper, we concentrate on the $HW^+W^-$ vertex, and study its higher order perturbative corrections within the SM.

To setup the context, we will consider the decay process $H \to W^*W \to \nu_e e W$ in this paper, although our results can also be applied to the production process $e^+e^- \to H\nu_e\bar{\nu}_e$. The leading order (LO) results \cite{Rizzo:1980gz, Pocsik:1980ta, Keung:1984hn} and the next-to-leading order (NLO) electroweak (EW) corrections \cite{Fleischer:1980ub} for this decay process have been known many years ago. Beyond the NLO, contributions enhanced by the large top quark mass have been considered at two-loops \cite{Kniehl:1995ra, Djouadi:1997rj} and three-loops \cite{Frink:1996sv}, albeit the assumption $m_H > 2m_W$. For a light Higgs boson, Ref.~\cite{Kniehl:2012rz} reconsidered these calculations, and also gave predictions for the differential decay rates.

In this paper, we provide exact analytic results for the next-to-next-to-leading order (NNLO) mixed QCD-EW corrections to the $HW^+W^-$ vertex at order $\alpha\alpha_s$. The analytic expressions are valid for arbitrary values of the external momenta. Hence it is applicable to cases where both $W$ bosons are off-shell, and also to cases where the large top mass approximation may not hold. We apply these analytic results to the Higgs decay process, and provide numeric predictions for both integrated and differential decay rates. These results will be useful for extracting the $HW^+W^-$ coupling from future high precision experimental data.

This paper is organized as follows. In Section~\ref{sec:notation} we briefly introduce our notations and the LO and NLO results. In Section~\ref{sec:nnlo} we present the analytic calculation of the mixed QCD-EW corrections for the $HW^+W^-$ vertex, with generic momenta configurations. In Section~\ref{sec:decay}, we apply our analytic results to the $H \to \nu_e e W$ decay process, and obtain numeric predictions. The summary and outlook come in Section~\ref{sec:summary}, and we leave some lengthy expressions to the Appendix.

\section{Notations and lower-order results}
\label{sec:notation}

In this section we setup our notations and present the LO results as well as the NLO EW corrections for the partial decay width. In our calculation we will neglect masses of all light fermions except that of the top quark. At LO we assign the momenta as
\begin{equation}
H(P) \to W_{\nu}^{*\pm}(q) + W_{\mu}^\mp(p_3) \to \overset{\scriptscriptstyle(-)}{\nu_e}(p_1) + e^\pm(p_2) + W_{\mu}^\mp(p_3) \, .
\label{eq:momenta}
\end{equation}
The scalar products of the momenta lead to
\begin{align}
&P^2 = m_H^2 \, , \quad p_1^2=p_2^2=0 \, , \quad p_3^2 = m_W^2 \, ,\notag \\
&q^2 = (p_1+p_2)^2 = Q^2 = M_{e\nu}^2 \, , \quad (p_2+p_3)^2 = M_{e W}^2 \, .
\end{align}
The LO partial decay width is then written as
\begin{equation}
\Gamma_{\text{LO}} = \sum_{\text{spin,color}} \int d\text{PS}_3 \, \frac{1}{2m_H} \left| M_{\text{LO}} \right|^2 ,
\label{eq:GammaLO}
\end{equation}
where the 3-body phase-space integration measure is given by
\begin{equation}
d\text{PS}_3 = \left( \prod_{i=1}^3 \frac{d^3p_i}{(2\pi)^3}\frac{1}{2E_{i}} \right) (2\pi)^4 \, \delta^{(4)}(P-p_1-p_2-p_3) \, .
\label{eq:PS3}
\end{equation}


In general the 3-body phase-space integral can be performed numerically with a Monte Carlo program, which will be useful when we compute differential decay widths. For the fully integrated decay width at LO, the phase space integral can be calculated analytically and we find
\begin{multline}
\Gamma_{\text{LO}} = \frac{\alpha^2 m_H}{384 \pi (2r)^4 s_W^4} \bigg[ (1-4r^2)(47-52r^2+32r^4) + 24 (1-6r^2+4r^4) \log(2r)
\\
+\frac{24 r(5-8r^2+4r^4) \cos^{-1}\big(r(3-4r^2)\big)}{\sqrt{1-r^2}} \bigg] \, ,
\end{multline}
where $s_W = \sin\theta_W$ and $r=m_H/(2m_W)$.
The NLO EW corrections to this decay process involve closed fermion loops, exchanges of electroweak gauge bosons and Higgs boson, as well as real photon emissions. For these corrections we invoke the program \texttt{MadGraph5\_aMC@NLO} \cite{Frederix:2018nkq} which can automatically compute NLO QCD and EW corrections to standard model processes. In the rest of the paper, we present our calculation of the NNLO mixed QCD-EW corrections at order $\alpha \alpha_s$.

\section{The $HW^+W^-$ vertex at two loops}
\label{sec:nnlo}

\begin{figure}[t!]
	\centering
	\includegraphics[width=0.9\textwidth]{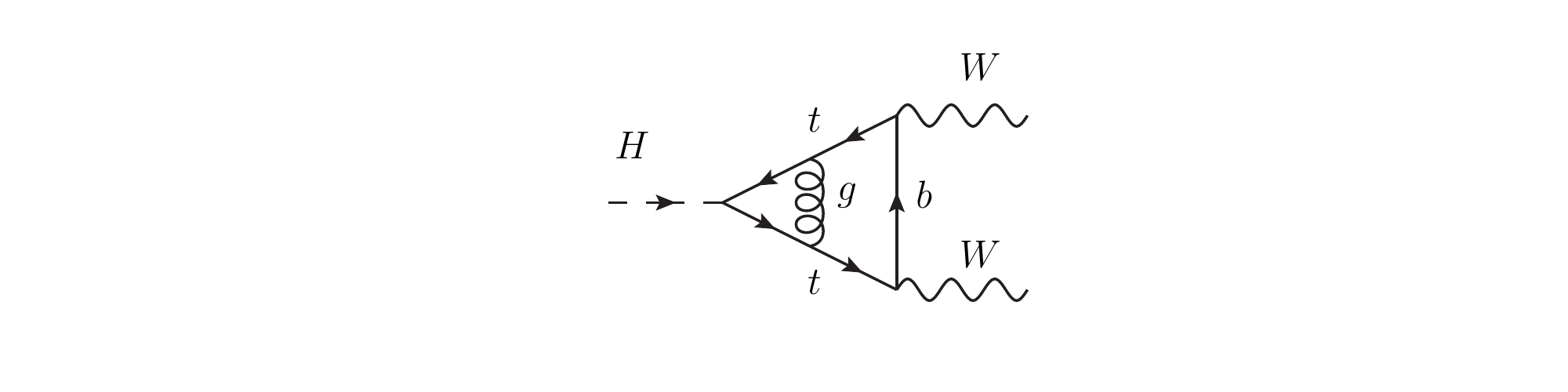}
	\vspace{-3ex}
	\caption{\label{figure:diagram}A typical Feynman diagram contributing to the mixed QCD-EW corrections for the $HW^+W^-$ vertex.}
\end{figure}

In this section, we present the calculation of the two-loop form factors entering the mixed QCD-EW corrections for the $HW^+W^-$ vertex. This serves as a main new result of this paper. A typical Feynman diagram is depicted in Fig.~\ref{figure:diagram}. The most generic Lorentz structure of the $HW^+W^-$ vertex can be written as:
\begin{equation}
\mathcal{T}^{\mu\nu} =\frac{i e m_W}{s_W} (g^{\mu\nu} T_1 + p^{\mu}_3 p^{\nu}_3 T_2 + p^{\mu}_3 q^{\nu} T_3 + p^{\nu}_3 q^{\mu} T_4 + q^{\mu} q^{\nu} T_5) \, .
\label{eq:tensor}
\end{equation}
Among the scalar coefficient functions $T_i$ in the above formula, only $T_1$ and $T_4$ contribute to the partial decay width. At NNLO, the coefficient functions can be written as linear combinations of scalar two-loop integrals of the form
\begin{equation}
F_{\{a_i\}} \equiv m_t^{4\epsilon} \int \frac{d^dk_1}{i\pi^{d/2}\Gamma(\epsilon)} \frac{d^dk_2}{i\pi^{d/2}\Gamma(\epsilon)} \prod_{i=1}^7 \frac{1}{D_i^{a_i}} \, ,
\label{eq:integrals}
\end{equation}
where $d=4-2\epsilon$ in dimensional regularization, $k_1$ and $k_2$ are loop momenta, and $\{a_i\} = \{a_1,a_2,a_3,a_4,a_5,a_6,a_7\}$ are the powers of propagator denominators $D_i$ in the integrand. The propagator denominators are given by
\begin{align*}
\{& k_1^2-m_t^2, \, (k_1-p_3-q)^2-m_t^2, \, k_2^2-m_t^2, \, (k_2-p_3-q)^2-m_t^2,\\
&(k_1-q)^2, \, (k_2-q)^2, \, (k_1-k_2)^2 \} \, .
\end{align*}
In practice, these denominators cannot simultaneously appear with positive powers in a Feynman diagram. Nevertheless, for completeness we will consider the top level topology $F_{1,1,1,1,1,1,1}$ and all its sub-topologies. We perform the integration-by-parts (IBP) reduction with the program packages \texttt{FIRE6} \cite{Smirnov:2019qkx} and \texttt{LiteRed} \cite{Lee:2013mka}, and cross-check using \texttt{Kira} \cite{Klappert:2020nbg}. We find 38 master integrals in total. We note that some of the sub-topologies have been studied in \cite{DiVita:2017xlr}, with 31 master integrals.\footnote{After our current paper, another work \cite{Janowski:2021yvz} appeared which also calculated these integrals.} The additional master integrals we get are:
\begin{align*}
\{ F_{0,1,0,1,1,1,0}, \, F_{0,1,1,1,1,0,1}, \, F_{0,1,1,1,1,0,2}, \, F_{0,1,2,1,1,0,1}, \, F_{0,2,1,1,1,0,1}, \, F_{0,1,1,1,1,1,0}, \, F_{1,1,1,1,1,1,0} \} \, .
\end{align*}
The last integral in the above does not appear in the amplitude, and is included because we perform the reduction from the top topology with 7 propagators. The other six integrals are required in the calculation of the two-loop amplitude. They can be related to $\mathcal{T}_{11}$, $\mathcal{T}_{25}$, $\mathcal{T}_{27}$--$\mathcal{T}_{30}$ of Figure 4 in \cite{DiVita:2017xlr} by an exchange of external momenta $p_1 \leftrightarrow p_2$ ($p_3 \leftrightarrow q$ in our notation).

To calculate the master integrals, we will employ the method of canonical differential equations \cite{Kotikov:1990kg, Kotikov:1991pm,Remiddi:1997ny,Gehrmann:1999as,Argeri:2007up,MullerStach:2012mp,Henn:2013pwa,Henn:2014qga,Ablinger:2015tua,Adams:2017tga,Bosma:2017hrk}. For later convenience, we introduce the following dimensionless variables:
\begin{align}
u=-\frac{m_H^2}{4m_t^2} \, , \quad v=-\frac{m_W^2}{4m_t^2} \, , \quad w=-\frac{Q^2}{4m_t^2} \, .
\label{eq:uvw}
\end{align}

\subsection{Canonical basis of the two-loop integrals}

We would like to find linear combinations of integrals as in Eq.~\eqref{eq:integrals} which satisfy the canonical-form differential equations. Namely, the canonical basis $\vec{f}(u,v,w;\epsilon)$ of the master integrals satisfy differential equations of the form
\begin{align}
d\vec{f}(u,v,w;\epsilon) &= \epsilon \, dA(u,v,w) \, \vec{f}(u,v,w;\epsilon) \nonumber
\\
&= \epsilon \sum_{i} A_i \, d\log(\alpha_i) \, \vec{f}(u,v,w;\epsilon) \, ,
\label{eq:deq1}
\end{align}
where $A_i$ are constant matrices independent of kinematic variables and the dimensional regulator. The ``letters'' $\alpha_i \equiv \alpha_i(u,v,w)$ are algebraic functions of the kinematic variables $u$, $v$ and $w$.

To find the canonical basis $\vec{f}(u,v,w;\epsilon)$, we use the method of \cite{Chen:2020uyk} supplemented by suitable linear transformations to further simplify the differential equations. The method of \cite{Chen:2020uyk} utilizes the Baikov representation \cite{Baikov:1996iu} and look for $d\log$-form integrals which serve as candidates for the canonical basis. The Baikov representation for integrals of the form Eq.~\eqref{eq:integrals} can be written as
\begin{equation}
F_{\{a_i\}} = \mathcal{N}_\epsilon \int_{\mathcal{C}}  u(\bm{z}) \, \frac{d^7\bm{z}}{z_1^{a_1} z_2^{a_2} z_3^{a_3} z_4^{a_4} z_5^{a_5} z_6^{a_6} z_7^{a_7}} \, ,
\label{eq:baikov}
\end{equation}
where $z_i \equiv D_i$ are the propagator denominators which become integration variables in the Baikov representation, and the vector $\bm{z}$ is the collection of $z_i$. The prefactor $\mathcal{N}_\epsilon$ and the function $u(\bm{z})$ are given by
\begin{equation}
\mathcal{N}_\epsilon = \frac{m_t^{4\epsilon}\lambda^{-1/2+\epsilon}}{4\pi^3\Gamma(1-2\epsilon)\Gamma^2(\epsilon)} \, , \quad u(\bm{z}) = \big[ P_0(\bm{z}) \big]^{-1/2-\epsilon} \, , \quad P_0(\bm{z}) = G(k_1,k_2,p_3,q) \, ,
\end{equation}
where $\lambda \equiv \lambda(m_H^2,m_W^2,Q^2)$, with $\lambda(x,y,z) \equiv x^2 + y^2 + z^2 - 2x y - 2y z - 2z x$ being the K\"all\'en function. The Gram determinant is defined by
\begin{equation}
G(q_1,\ldots,q_n) \equiv \det (-q_i \cdot q_j) \equiv \det
\begin{pmatrix}
-q_1 \cdot q_1 & -q_1 \cdot q_2 & \cdots & -q_1 \cdot q_n
\\
-q_2 \cdot q_1 & -q_2 \cdot q_2 & & \vdots
\\
\vdots & & \ddots & \vdots
\\
-q_n \cdot q_1 & \cdots & \cdots & -q_n \cdot q_n
\end{pmatrix}
\, .
\end{equation}
The integration domain $\mathcal{C}$ in Eq.~\eqref{eq:baikov} has the property that $u(\bm{z})$ vanishes on its boundary $\partial\mathcal{C}$.

For illustration purposes, we consider the $\{1,0,1,1,1,0,1\}$ sub-topology, which turns out to be the most complicated one. This includes all integrals of the form Eq.~\eqref{eq:baikov} with the powers $a_1$, $a_3$, $a_4$, $a_5$ and $a_7$ being positive, and the powers $a_2$ and $a_6$ being non-positive. For the construction of $d\log$-form integrals, it is reasonable to first consider the cases where either $a_2$ or $a_6$ is fixed to zero. As an example, we will fix $a_2 = 0$. It is then possible to perform the integration over $z_2$ in Eq.~\eqref{eq:baikov}, resulting in
\begin{equation}
\mathcal{N}_\epsilon \int_{\mathcal{C}} u(\bm{z}) \, dz_2 \equiv \mathcal{N}'_{\epsilon} \, u_2(\bm{z}') = \mathcal{N}'_{\epsilon} \, P_1^{-1/2+\epsilon} \, P_2^{-\epsilon} \, P_3^{-\epsilon} \, , 
\end{equation}
where $\bm{z}'$ is given by $\bm{z}$ with $z_2$ removed. The new prefactor and the polynomials are given by
\begin{align}
\mathcal{N}'_{\epsilon} &= \frac{m_t^{4\epsilon}\lambda^{-1/2+\epsilon}}{4\pi^2\Gamma^2(1-\epsilon)\Gamma^2(\epsilon)} \, , \nonumber
\\
P_1 &= P_1(z_3, z_6) = -4G(k_2, q) \, , \nonumber
\\ 
P_2 &= P_2(z_1, z_3, z_5, z_7, z_6) = 4G(k_1, k_2, q) \, , \nonumber
\\
P_3 &= P_3(z_3, z_4, z_6) = 4G(k_2, p_3, q) \, .
\end{align}
The integrals in this sub-topology with $a_2=0$ can then be expressed as
\begin{equation}
F_{\{a_i\}} = \mathcal{N}'_\epsilon \int_{\mathcal{C}}  u_2(\bm{z}') \, \frac{d^6\bm{z}'}{z_1^{a_1} z_3^{a_3} z_4^{a_4} z_5^{a_5} z_6^{a_6} z_7^{a_7}} \, ,
\label{eq:baikov2}
\end{equation}
which is equivalent to a loop-by-loop Baikov representation.

We now want to search for rational functions $\hat{\varphi}(\bm{z}')$ such that the integrand $u_2(\bm{z}') \hat{\varphi}(\bm{z}')$ takes the $d$-dimensional generalized $d\log$-form \cite{Chen:2020uyk}. We perform the construction in the variable-by-variable approach in the order $z_6, z_1, z_5, z_3, z_4, z_7$, and find 4 candidates for the canonical Feynman integrals in this sub-topology:
\begin{align}
\hat{\varphi}_1 &= \frac{\sqrt{\lambda}}{z_1z_3z_4z_5z_7} \, , \nonumber
\\
\hat{\varphi}_2 &= \frac{\sqrt{\lambda}\sqrt{m_H^2(m_H^2-4m_t^2)}}{z_1z_3z_4z_5z_7} \frac{1}{P_{3}}\frac{\partial P_{3}}{\partial z_4} \, , \nonumber
\\
\hat{\varphi}_3 &= \frac{1}{z_1z_3z_4z_5z_7} \left[ \frac{1}{P_{3}} \frac{\partial P_{3}}{\partial z_4} \frac{\partial P_{3}}{\partial z_6} - 2 \frac{\partial^2 P_{3}}{\partial z_4 \partial z_6} \right] , \nonumber
\\
\hat{\varphi}_4 &= \frac{\sqrt{\lambda}(Q^2 -m_t^2)}{z_1z_3z_4z_5z_7} \frac{1}{P_{2}}\frac{\partial P_{2}}{\partial z_5} \, .
\end{align}
These candidates can be converted to Feynman integrals of the form \eqref{eq:baikov} using intersection theory \cite{Mizera:2017rqa, Mastrolia:2018uzb, Frellesvig:2019kgj, Frellesvig:2019uqt, Mizera:2019vvs, Weinzierl:2020xyy, Frellesvig:2020qot} or via IBP relations. The results are given by
\begin{align}
\varphi_1 &= -\frac{16 m_t^4 R_1 R_2}{\epsilon } F_{1,0,1,2,1,0,1} \, , \nonumber
\\
\varphi_2 &= -\frac{4  m_t^2 (u+v-w)}{\epsilon } F_{0,2,1,0,0,1,1} -8  m_t^2 (u-v+w) F_{1,0,1,1,1,0,1} \nonumber \\
&\phantom{={}} -\frac{8 m_t^4 u (2 u-2 v-2 w+1)}{\epsilon } F_{1,0,1,2,1,0,1} +\frac{8 m_t^2 u}{\epsilon } F_{1,0,1,2,1,-1,1} \, , \nonumber
\\
\varphi_3 &= \frac{4 m_t^4 (4 w+1) R_2}{\epsilon} F_{1,0,1,1,2,0,1} \, , \nonumber
\\
\varphi_4 &= -\frac{16 m_t^4 w R_2}{\epsilon} F_{2,0,1,1,1,1,0} - \frac{2 m_t^2 R_2}{\epsilon} F_{0,2,1,1,0,1,0} \, .
\end{align}
Where we have introduced the following two square roots:
\begin{equation}
R_1 \equiv R_1(u) = \sqrt{u(u+1)} \, , \quad R_2 \equiv R_2(u,v,w) = \sqrt{\lambda(u,v,w)} \, .
\label{eq:R1R2}
\end{equation}
Applying the same procedure to other (sub)-topologies, we are able to construct 38 $d\log$-form integrals in the Baikov representation, and convert them to Feynman integrals using either traditional IBP method or the intersection theory. We have checked that they indeed satisfy the canonical-form differential equations \eqref{eq:deq1}. Further (simple) linear transforms can be applied to simplify the matrices $A_i$. The final form of the canonical basis $\vec{f}(u,v,w;\epsilon)$ in terms of linear combinations of Feynman integrals is given in Appendix~\ref{Appendix:basis}.

\subsection{Solution to the canonical differential equations}

The differential equations \eqref{eq:deq1} involve 18 independent letters as following:
\begin{align}
\bigg\{ &u, \frac{R_1-u}{R_1+u}, u+1, v, 4v+1, w, 4w+1, \lambda(u,v,w), \nonumber
\\
&\frac{v-w-R_1-R_2}{v-w+R_1-R_2}, \frac{v-w+R_1-R_2}{v-w-R_1+R_2}, \frac{v-w-R_1-R_2}{v-w+R_1+R_2}, \nonumber
\\
&\frac{u+v-w-R_2}{u+v-w+R_2},\frac{u-v+w-R_2}{u-v+w+R_2}, \frac{2R_2+g(u,v,w)}{2R_2-g(u,v,w)}, \frac{2R_2+g(u,w,v)}{2R_2-g(u,w,v)}, \nonumber
\\
&h(u,v,w), h(u,w,v), 4 (v-w)^2-u (4 v+1) (4 w+1) \bigg\} \, ,
\end{align}
where we have introduced the following polynomials:
\begin{align}
g(u,v,w) &= 1 + 2u + 2v - 2w \, , \nonumber
\\
h(u,v,w) &= 1 + 4u + 4v + 16uv - 4w \, ,
\label{equ:lgh}
\end{align}
The solution to the canonical form differential equations \eqref{eq:deq1} can be formally written as a path-ordered exponential integrated from the boundary point $\vec{x}_0=(u_0,v_0,w_0)$ to the point $\vec{x}=(u,v,w)$
\begin{equation}
\vec{f}(u,v,w;\epsilon)=\mathcal{P} \exp \left(\epsilon \int_{\vec{x}_0}^{\vec{x}} dA \right)\vec{f}(u_0,v_0,w_0;\epsilon) \, .
\label{eq:exponential}
\end{equation}
We choose the boundary point to be $\vec{x}_0=(0,0,0)$, which corresponds to zero external momenta. The values of the mater integrals at the boundary can be easily obtained and are given by
\begin{align}
f_1(0,0,0;\epsilon) &= 1 \, , \nonumber
\\
f_5(0,0,0;\epsilon) &= f_7(0,0,0;\epsilon) = -\frac{\Gamma(1-\epsilon ) \Gamma (1+2\epsilon)}{\Gamma(1+\epsilon)} \, , \nonumber
\\
f_{22}(0,0,0;\epsilon) &= f_{25}(0,0,0;\epsilon) = \frac{1}{8} - \frac{\Gamma(1-\epsilon ) \Gamma(2\epsilon)}{4\Gamma(\epsilon)} \, ,
\end{align}
with $f_i(0,0,0;\epsilon) = 0$ for all other $i$'s.

The path-ordered exponential \eqref{eq:exponential} can be expanded as power series in $\epsilon$. At each order in $\epsilon$ we need to evaluate iterated integrals in terms of generalized polylogarithms (GPLs) \cite{Goncharov:1998kja}. For that it is convenient to perform a change of variables such that the integration kernels are rational functions without square roots. There are two ways to achieve that, which lead to the same final results. The first is to perform variable changes on $u$, $v$ and $w$, such that both $R_1$ and $R_2$ become rational functions of the new variables. This can be done using the method of \cite{Besier:2018jen, Besier:2019kco}. The second way, which we'll describe in more details, amounts to choosing a specific integration path in the $(u,v,w)$ space, such that the integration reduces to a single-variable problem. Observing that $R_2$ is homogeneous in $u$, $v$ and $w$, it is reasonable to pick a straight path from $(0,0,0)$ to $(u,v,w)$ parameterized by a variable $t$ as $(tu,tv,tw)$. During the integration over $t$, $R_2$ simply becomes $t$ multiplied by a constant factor:
\begin{equation}
R_2(tu,tv,tw) = t R_2(u,v,w) \, .
\end{equation}
On the other hand, we now need to perform a variable change on $t$ such that $R_1(tu)$ becomes a rational function. This can be simply done with
\begin{equation}
t = \frac{x^2}{4u(1-x)} \, , \quad R_1(tu) = \frac{x(x-2)}{4 (x-1)}\, .
\end{equation}
When $t$ goes from $0$ to $1$, $x$ goes from $0$ to $2(R_1(u)-u)$. This determines the integration domain over $x$.

After the variable change, the iterated integrals at order $\epsilon^n$ have the generic structure
\begin{equation}
\int dx \, \frac{1}{x - x_i} G^{(n-1)}(x) \, ,
\end{equation}
with $G^{(n-1)}(x)$ being functions at order $\epsilon^{n-1}$. These integrals can be naturally evaluated in terms of GPLs, with the branch points $x_i$ given by
\begin{align}
x_1 &= 1 \, , \quad x_2 = 2 \, , \quad x_3 = 0 \, , \quad x_4 = \frac{u+v-w+R_2}{2v} \, , \quad x_5 = \frac{u+v-w-R_2}{2v} \, , \nonumber
\\
x_6 &= \frac{u-v+w+R_2}{2w} \, , \quad x_7 = \frac{u-v+w-R_2}{2w} \, , \quad x_8 = \frac{u-\sqrt{u(u-4v)}}{2v} \, , \nonumber
\\
x_9 &= \frac{u+\sqrt{u(u-4v)}}{2v} \, , \quad x_{10} = \frac{u-\sqrt{u(u-4w)}}{2w} \, , \quad x_{11} = \frac{u+\sqrt{u(u-4w)}}{2w} \, , \nonumber
\\
x_{12} &= \frac{u + \sqrt{u (2v-2w-u+2R_2)}}{u-v+w-R_2} \, , \quad x_{13} = \frac{u - \sqrt{u (2v-2w-u+2R_2)}}{u-v+w-R_2} \, , \nonumber
\\
x_{14} &= \frac{u + \sqrt{u (2w-2v-u+2R_2)}}{u+v-w-R_2} \, , \quad x_{15} = \frac{u - \sqrt{u (2w-2v-u+2R_2)}}{u+v-w-R_2} \, , \nonumber
\\
x_{16} &= \frac{u - \sqrt{u (2w-2v-u-2R_2)}}{u+v-w+R_2} \, , \quad x_{17} = \frac{u + \sqrt{u (2w-2v-u-2R_2)}}{u+v-w+R_2}, \nonumber
\\
x_{18} &= \frac{u - \sqrt{u (2v-2w-u-2R_2)}}{u-v+w+R_2} \, , \quad x_{19} = \frac{u + \sqrt{u (2v-2w-u-2R_2)}}{u-v+w+R_2} \, .
\end{align}
The resulting GPLs can be evaluated numerically using program libraries \texttt{GiNaC} \cite{Bauer:2000cp, Vollinga:2005pk, Vollinga:2004sn} and \texttt{handyG} \cite{Naterop:2019xaf}. We have checked that our results for the master integrals are in good agreement with the numeric results of \texttt{pySecDec} \cite{Borowka:2017idc}.

\section{Application to the decay process}
\label{sec:decay}

\subsection{Analytic results at NNLO}

In this section, we apply the two-loop integrals calculated in the previous section to the decay process $H \to e \nu_e W$ at order $\alpha\alpha_s$. The decay process receives three kinds of contributions at this order: the corrections to the $HW^+W^-$ vertex, the corrections to the propagator of the off-shell $W^\pm$ field, and the corrections to the $e\nu_eW$ vertex (arising from counter-terms). In practice these can all be described by their contributions to the coefficient functions in Eq.~\eqref{eq:tensor} (with the third kind of contributions only affecting $T_1$). The squared amplitude at this order can then be written as
\begin{align}
2\Re(M^*_{\text{LO}} M_{\alpha\alpha_s}) &= \frac{8 \pi^2 \alpha^2}{s_W^4 \left( M_{e\nu}^2 - m_W^2 \right)^2} \bigg[ \Re(T_4) \left( M_{e\nu}^2-m_H^2+m_W^2 \right) \nonumber
\\
&\hspace{8em} \times \left( M_{e\nu}^2 M_{e W}^2 + (M_{e W}^2-m_H^2)(M_{e W}^2-m_W^2) \right) \nonumber
\\
&- 2\Re(T_1) \left( M_{e\nu}^2 (M_{e W}^2-2m_W^2)+(M_{e W}^2-m_H^2) (M_{e W}^2-m_W^2) \right) \bigg] \, .
\end{align}

To calculate the coefficient functions, we generate the relevant Feynman diagrams using \texttt{FeynArts} \cite{Hahn:2000kx}. The resulting amplitudes are further manipulated with \texttt{Mathematica}. In addition to the Feynman integrals calculated in the previous section, we also need to calculate a couple of extra integrals arising in the $W^\pm$ propagator and in the counter-terms. We have checked these results against those in the literature \cite{Djouadi:1993ss, Dittmaier:2015rxo}.

We renormalize the fields and the masses in the on-shell scheme, with pole masses $m_W$, $m_Z$, $m_H$ and $m_t$ as input parameters. The weak mixing angle is defined by the on-shell relation $c_W \equiv \cos(\theta_W) = m_W/m_Z$, whose renormalization is given by $\delta c_W^2/c_W^2=\delta m_W^2/m_W^2-\delta m_Z^2/m_Z^2$.
For the renormalization of the fine structure constant $\alpha$, we choose two different schemes: the $\alpha(m_Z)$-scheme with $\alpha(m_Z)=\alpha(0)/(1-\Delta\alpha(m_Z))$ as the input parameter, where $\Delta\alpha(m_Z)$ encodes the contributions to the renormalization of $\alpha$ from light fermion loops (including leptons and 5 flavors of light quarks); and the $G_{\mu}$-scheme with the Fermi-constant $G_{\mu}$ determined from the muon decay data as the input parameter. In the $G_\mu$-scheme, the coupling is given by
\begin{equation}
\alpha_{G_{\mu}} = \frac{\sqrt{2}}{\pi} G_{\mu} m_W^2 \left( 1 - \frac{m_W^2}{m_Z^2} \right) .
\end{equation}

\subsection{Numeric results}

We now present the numeric predictions from our calculations. We choose the input parameters as $m_t=\unit{172.76}{\GeV}$, $m_H=\unit{125.1}{\GeV}$, $m_Z=\unit{91.1876}{\GeV}$, $m_W=\unit{80.379}{\GeV}$, $G_{\mu}=\unit{$1.1663787 \times 10^{-5}$}{\GeV^{-2}}$, $\alpha(m_Z)=1/128.929$ and $\alpha_s(m_Z)=0.1179$ \cite{Zyla:2020zbs}. The default renormalization scale for $\alpha_s$ is chosen as $\mu=m_H$.

In Table~\ref{tab1} and Table~\ref{tab2}, we show the LO, NLO EW, and NNLO QCD-EW predictions to the partial decay widths in the $\alpha(m_Z)$-scheme and $G_\mu$-scheme, respectively. We find that the mixed QCD-EW corrections amount to about 1\% of the LO prediction in the $\alpha(m_Z)$-scheme, and about 0.2\% of the LO prediction in the $G_\mu$-scheme. The smallness of the contributions in the $G_\mu$-scheme can be partly attributed to the fact that corrections to the $W$ propagator and to the $e\nu_eW$ vertex are absorbed into the coupling constant. Nevertheless, we find that the NNLO results in the two schemes are rather close to each other, showing the stability of our prediction against scheme changes.

\begin{table}[t!]
	\centering
	\begin{tabular}{|c|c|c|c|}
		\hline
		& LO & NLO EW & NNLO QCD-EW
		\\ \hline
		$\Gamma$ ($\unit{$10^{-5}$}{\GeV}$) & $4.597$ & $4.474$ & $4.518$
		\\ \hline
		$\delta\Gamma$ ($\unit{$10^{-5}$}{\GeV}$) & & $-0.123$ & $+0.044$
		\\ \hline
		$\delta\Gamma/\Gamma_\text{LO}$ & & $-2.7\%$ & $+1.0\%$
		\\ \hline
	\end{tabular}
	\caption{The partial decay widths at various orders in the $\alpha(m_Z)$ scheme.}\label{tab1}
\end{table}

\begin{table}[t!]
	\centering
	\begin{tabular}{|c|c|c|c|}
		\hline
		& LO & NLO EW & NNLO QCD-EW
		\\ \hline
		$\Gamma$ ($\unit{$10^{-5}$}{\GeV}$) & $4.374$ & $4.524$ & $4.531$
		\\ \hline
		$\delta\Gamma$ ($\unit{$10^{-5}$}{\GeV}$) & & $+0.150$ & $+0.007$
		\\ \hline
		$\delta\Gamma/\Gamma_\text{LO}$ & & $+3.4\%$ & $+0.2\%$
		\\ \hline
	\end{tabular}
	\caption{The partial decay widths at various orders in the $G_{\mu}$ scheme.}\label{tab2}
\end{table}

\begin{figure}[t!]
	\centering
	\includegraphics[width=0.48\textwidth]{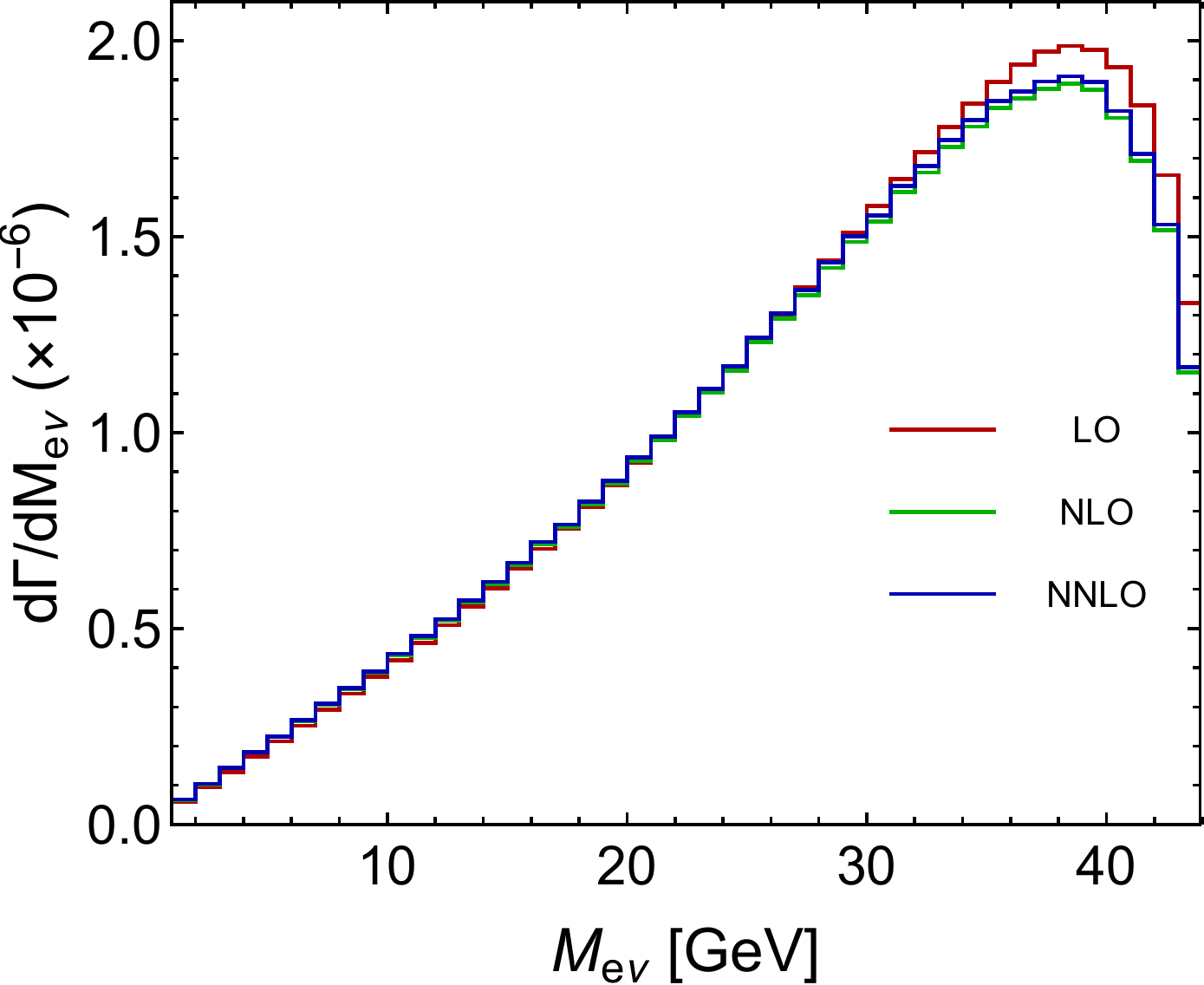}
	\includegraphics[width=0.48\textwidth]{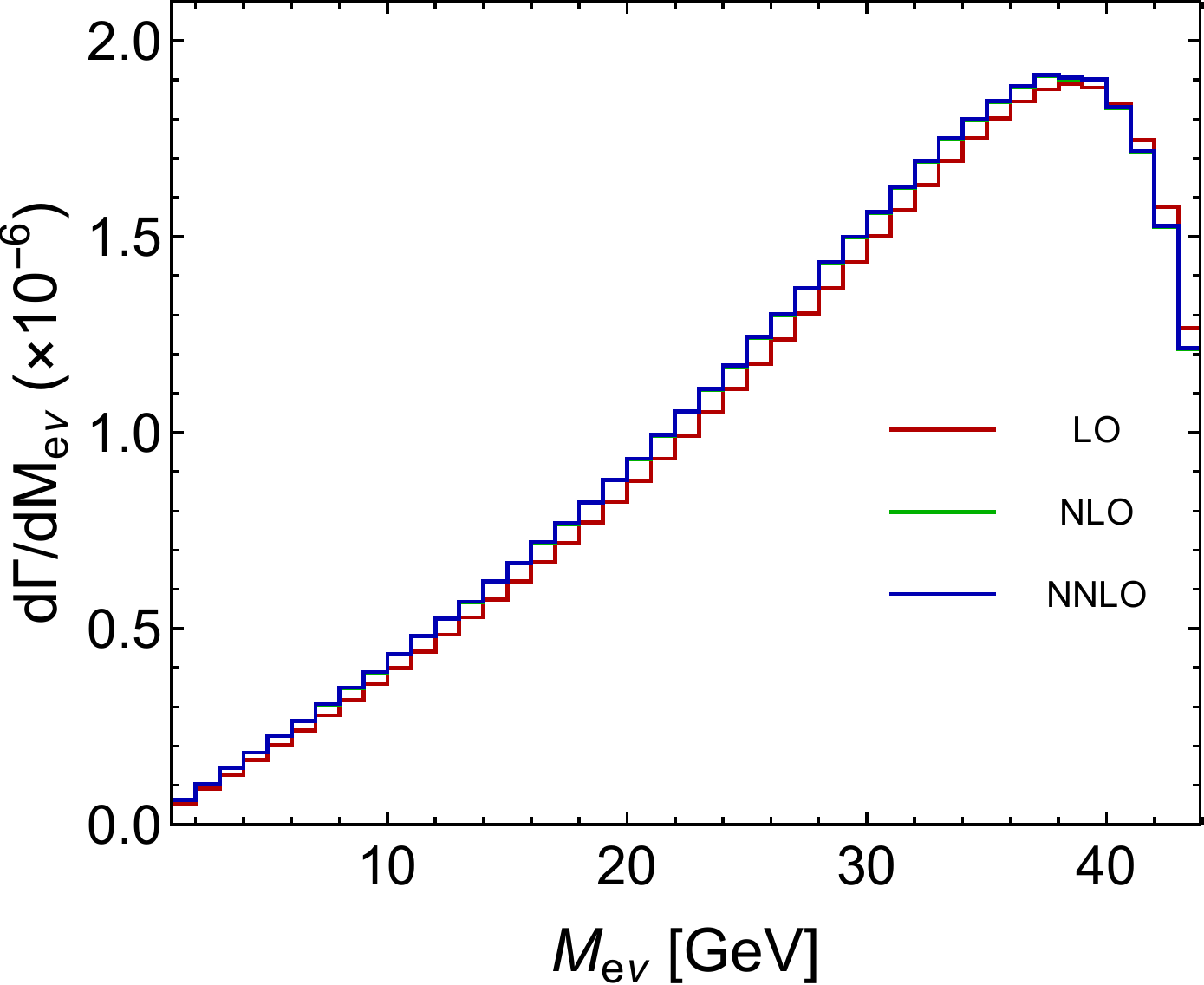}
	\vspace{1ex}
	\caption{\label{fig:dGamma1}The $M_{e\nu}$ distribution in the $\alpha(m_Z)$ scheme (left) and in the $G_{\mu}$ scheme (right).}
\end{figure}

\begin{figure}[t!]
	\centering
	\includegraphics[width=0.48\textwidth]{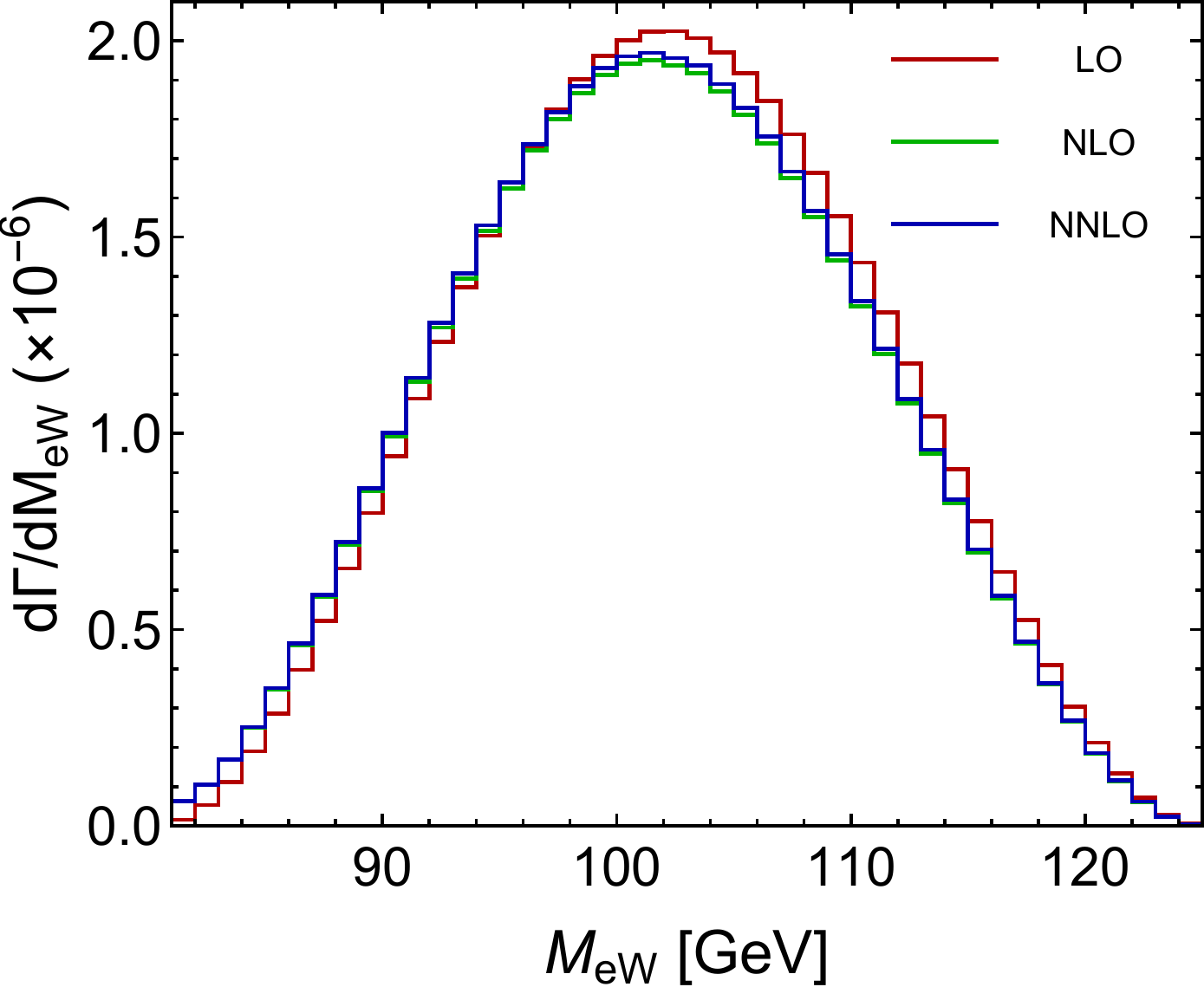}
	\includegraphics[width=0.48\textwidth]{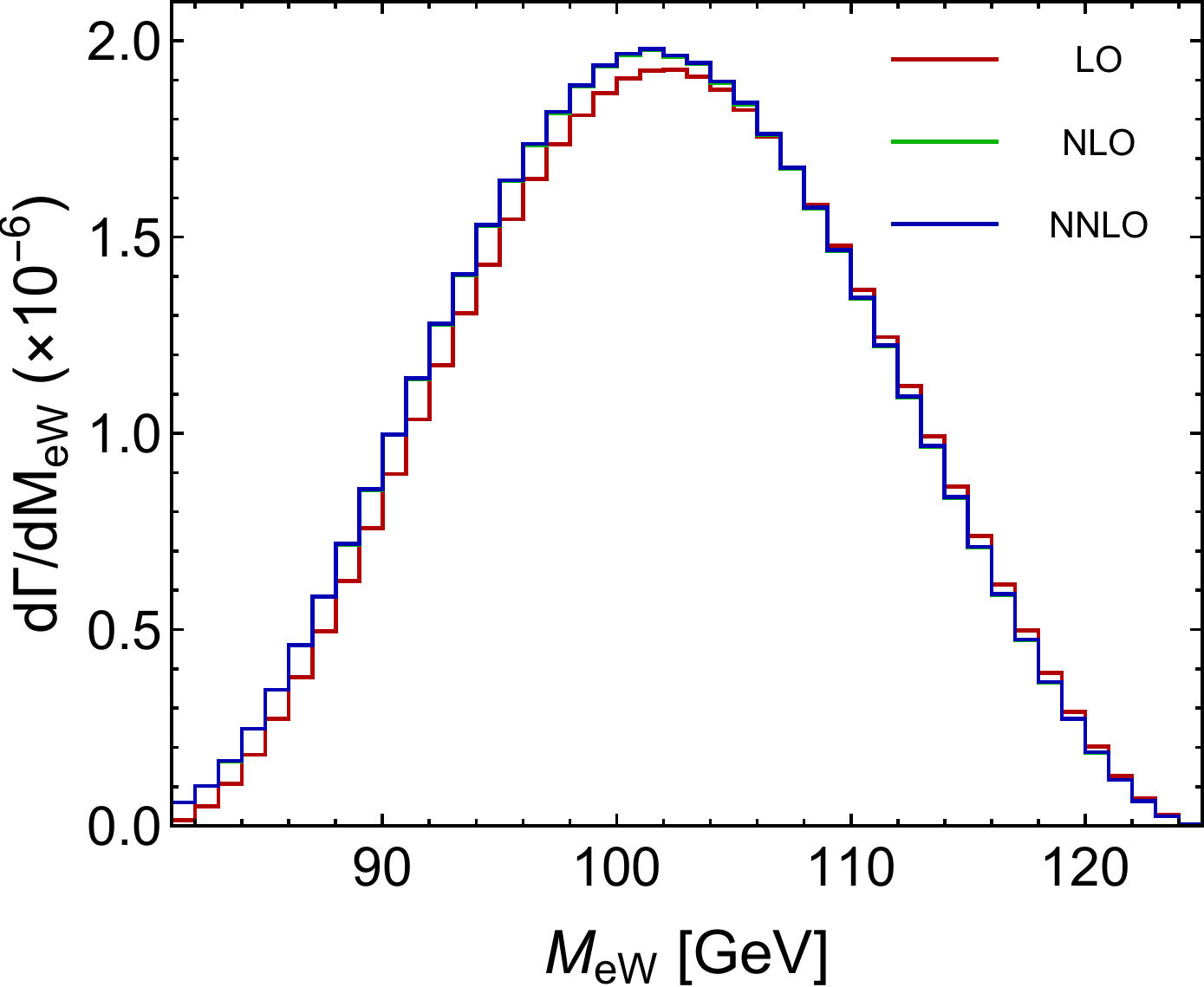}
	\vspace{1ex}
	\caption{\label{fig:dGamma2}The $M_{eW}$ distribution in the $\alpha(m_Z)$ scheme (left) and in the $G_{\mu}$ scheme (right).}
\end{figure}

We now proceed to study the differential decay rates. In Fig.~\ref{fig:dGamma1} we depict the $M_{e\nu}$ distribution at various orders in perturbation theory, in the two renormalization schemes mentioned above. This distribution gives us information about the virtuality of the off-shell $W$ boson. Here we observe similar behaviors as for the integrated decay rate. The NNLO corrections are much smaller in the $G_\mu$ scheme, as shown in the right plot in Fig.~\ref{fig:dGamma1}, where the green and blue curves almost completely overlap with each other. Since the neutrino is not observable (although could be reconstructed), it is also interesting to study the invariant mass of the visible part of the decay products. In Fig.~\ref{fig:dGamma2} we show the $M_{eW}$ distribution, again in the two schemes at different orders. This distribution can be measured at the LHC and the future Higgs factories, with the $W$ boson decaying hadronically. Our results then provide the high precision theoretical predictions to be compared with the experimental data.

\section{Summary and Outlook}
\label{sec:summary}

In this paper, we have studied a class of two-loop triangle integrals entering the $\mathcal{O}(\alpha\alpha_s)$ corrections to the $HW^+W^-$ vertex. We have constructed a canonical basis consisting of 38 master integrals using the Baikov representation and intersection theory. We have derived the $\epsilon$-form differential equations for the master integrals, and are able to find fully analytic solutions in terms of GPLs. The fully analytic form allows fast and accurate numeric evaluation for any combination of external momenta and internal masses, which is important for phenomenological studies.

We apply our results to the $H \rightarrow \nu_e e W$ decay process. For the integrated decay width, we find that the NNLO mixed QCD-EW corrections can reach 1\% of the LO result in the $\alpha(m_Z)$ scheme. The size of the corrections are reduced to about 0.2\% in the $G_{\mu}$ scheme. This seems to indicate that the perturbative convergence in the $G_{\mu}$ scheme is better. However, this needs to be confirmed by the behavior of the purely electroweak two-loop corrections of order $\alpha^2$. We have also studied the differential decay rates and drawn similar conclusions.

For the decay process, the complete $\mathcal{O}(\alpha\alpha_s)$ corrections are not expected to be quite different in size compared to those calculated in the heavy top limit \cite{Kniehl:1995ra, Djouadi:1997rj, Frink:1996sv, Kniehl:2012rz}. However, the fully analytic form of our results enable us to study cases where the virtuality of $W^\pm$ bosons exceeds the top quark mass, where the heavy top approximation is no longer valid. This is the case for, e.g., the $W$ boson fusion process $W^{+*} W^{-*} \to H$. The $\mathcal{O}(\alpha\alpha_s)$ corrections for this production process will be presented in a forthcoming article.

\appendix

\section{The canonical basis for the master integrals}
\label{Appendix:basis}

In this appendix, we provide the canonical basis for the master integrals appearing in the $\mathcal{O}(\alpha\alpha_s)$ corrections to the $HW^+W^-$ vertex. The 38 integrals are given by
\begin{flalign*}
f_{1}&=F_{0,2,0,2,0,0,0},\\
f_{2}&=u m_t^2 F_{0,2,2,0,0,0,1},\\
f_{3}&=2 m_t^2 R_1 F_{0,2,1,0,0,0,2} +m_t^2 R_1 F_{0,2,2,0,0,0,1} ,\\
f_{4}&=v m_t^2 F_{0,0,0,2,2,0,1},\\
f_{5}&=(4 v+1)m_t^2 F_{0,0,0,1,2,0,2} +2m_t^2 F_{0,0,0,2,2,0,1} ,\\
f_{6}&=w m_t^2 F_{0,0,2,0,2,0,1},\\
f_{7}&=(4 w+1)m_t^2 F_{0,0,1,0,2,0,2} +2m_t^2 F_{0,0,2,0,2,0,1} ,\\
f_{8}&=m_t^2 R_1 F_{0,2,1,2,0,0,0},\\
f_{9}&=v m_t^2 F_{0,2,0,2,0,1,0},\\
f_{10}&=w m_t^2F_{0,2,2,0,0,1,0},\\
f_{11}&=m_t^4 R_1^2 F_{2,1,2,1,0,0,0},\\
f_{12}&=v^2 m_t^4 F_{0,2,0,2,1,1,0},\\
f_{13}&=w^2 m_t^4 F_{2,0,2,0,1,1,0},\\
f_{14}&=v w m_t^4 F_{0,2,2,0,1,1,0},\\
f_{15}&=w m_t^4 R_1 F_{2,0,2,1,1,0,0},\\
f_{16}&=v m_t^4 R_1 F_{0,2,2,1,1,0,0},\\
f_{17}&=\epsilon  m_t^2 R_2 F_{0,2,1,1,0,1,0},\\
f_{18}&=\epsilon  m_t^2 R_2 F_{0,0,1,1,2,0,1},\\
f_{19}&=m_t^2 R_1  (\epsilon  F_{0,0,1,1,2,0,1}+(2 \epsilon -1)
(F_{0,0,1,2,1,0,1}+F_{0,0,2,1,1,0,1})),\\
f_{20}&=\epsilon  m_t^2 R_2 F_{0,1,1,0,1,0,2},\\
f_{21}&=\epsilon  m_t^2 R_2 F_{0,1,2,0,1,0,1},\\
f_{22}&=-\frac{1}{4} m_t^2 (\epsilon (F_{0,1,1,0,1,0,2}+2 F_{0,1,2,0,1,0,1})
g(u,v,w)+F_{0,2,2,0,1,0,1} m_t^2 h(u,v,w)),\\
f_{23}&=\epsilon  m_t^2 R_2 F_{0,1,1,0,0,1,2},\\
f_{24}&=\epsilon  m_t^2 R_2 F_{0,2,1,0,0,1,1},\\
f_{25}&=-\frac{1}{4} m_t^2 (\epsilon (F_{0,1,1,0,0,1,2}+2 F_{0,2,1,0,0,1,1})
g(u,w,v)+F_{0,2,2,0,0,1,1} m_t^2 h(u,w,v)),\\
f_{26}&=\epsilon  m_t^4 R_1 R_2 F_{2,1,1,1,0,1,0},\\
f_{27}&=v \epsilon  m_t^4 R_2 F_{0,2,1,1,1,1,0},\\
f_{28}&=w \epsilon  m_t^4 R_2 F_{2,0,1,1,1,1,0},\\
f_{29}&=\epsilon ^2 m_t^2 R_2 F_{0,1,1,0,1,1,1},\\
f_{30}&=\epsilon ^2 m_t^2 R_2 F_{1,0,1,1,1,0,1},\\
f_{31}&=\epsilon  m_t^4 R_1 R_2 F_{1,0,1,2,1,0,1},\\
f_{32}&=-\epsilon ^2  m_t^2 (u-v+w)F_{1,0,1,1,1,0,1}- \epsilon   m_t^4 u (2 u-2v-2 w+1)F_{1,0,1,2,1,0,1}\\
&-\frac{1}{2} \epsilon   m_t^2 (u+v-w)F_{0,2,1,0,0,1,1}+ \epsilon m_t^2 u F_{1,0,1,2,1,-1,1},\\
f_{33}&=(4 w+1) \epsilon  m_t^4 R_2 F_{1,0,1,1,2,0,1},\\
f_{34}&=\epsilon ^2 m_t^2 R_2 F_{0,1,1,1,1,0,1},\\
f_{35}&=\epsilon  m_t^4 R_1 R_2 F_{0,1,2,1,1,0,1},\\
f_{36}&=-\epsilon ^2  m_t^2 (u+v-w)F_{0,1,1,1,1,0,1}- \epsilon m_t^4 u (2 u-2v-2 w+1)F_{0,1,2,1,1,0,1}\\
&-\frac{1}{2} \epsilon  m_t^2 (u-v+w) F_{0,1,2,0,1,0,1}+ \epsilon  m_t^2 u F_{0,1,2,1,1,-1,1},\\
f_{37}&=(4 v+1) \epsilon  m_t^4 R_2 F_{0,1,1,1,2,0,1},\\
f_{38}&=\epsilon ^2 m_t^4 R_2^2 F_{1,1,1,1,1,1,0} \, .
\end{flalign*}
where $u, v, w,$ $R_1$ and $R_2$ are defined by Eq.~\eqref{eq:uvw} and Eq.~\eqref{eq:R1R2}, respectively, and the functions $g(u,v,w)$ and $h(u,v,w)$ are given in Eq.~\eqref{equ:lgh}.

\bibliographystyle{JHEP}
\bibliography{cite}

\providecommand{\href}[2]{#2}\begingroup\raggedright\begin{thebibliography}{10}

\bibitem{Apollinari:2017cqg}
G.~Apollinari, O.~Br\"uning, T.~Nakamoto and L.~Rossi, \emph{{High Luminosity
  Large Hadron Collider HL-LHC}},
  \href{https://arxiv.org/abs/1705.08830}{{\ttfamily 1705.08830}}.

\bibitem{CEPCStudyGroup:2018ghi}
{\scshape CEPC Study Group} collaboration, \emph{{CEPC Conceptual Design
  Report: Volume 2 - Physics \& Detector}},
  \href{https://arxiv.org/abs/1811.10545}{{\ttfamily 1811.10545}}.

\bibitem{Abada:2019lih}
{\scshape FCC} collaboration, \emph{{FCC Physics Opportunities}: {Future
  Circular Collider Conceptual Design Report Volume 1}},
  \href{https://doi.org/10.1140/epjc/s10052-019-6904-3}{\emph{Eur. Phys. J. C}
  {\bfseries 79} (2019) 474}.

\bibitem{Bambade:2019fyw}
P.~Bambade et~al., \emph{{The International Linear Collider: A Global
  Project}},  \href{https://arxiv.org/abs/1903.01629}{{\ttfamily 1903.01629}}.

\bibitem{Sirunyan:2018koj}
{\scshape CMS} collaboration, \emph{{Combined measurements of Higgs boson
  couplings in proton\textendash{}proton collisions at $\sqrt{s}=13\,\text
  {Te}\text {V} $}},
  \href{https://doi.org/10.1140/epjc/s10052-019-6909-y}{\emph{Eur. Phys. J. C}
  {\bfseries 79} (2019) 421}
  [\href{https://arxiv.org/abs/1809.10733}{{\ttfamily 1809.10733}}].

\bibitem{Cepeda:2019klc}
M.~Cepeda et~al., \emph{{Report from Working Group 2}: {Higgs Physics at the
  HL-LHC and HE-LHC}},
  \href{https://doi.org/10.23731/CYRM-2019-007.221}{\emph{CERN Yellow Rep.
  Monogr.} {\bfseries 7} (2019) 221}
  [\href{https://arxiv.org/abs/1902.00134}{{\ttfamily 1902.00134}}].

\bibitem{Fleischer:1980ub}
J.~Fleischer and F.~Jegerlehner, \emph{{Radiative Corrections to Higgs Decays
  in the Extended Weinberg-Salam Model}},
  \href{https://doi.org/10.1103/PhysRevD.23.2001}{\emph{Phys. Rev. D}
  {\bfseries 23} (1981) 2001}.

\bibitem{Fleischer:1982af}
J.~Fleischer and F.~Jegerlehner, \emph{{Radiative Corrections to Higgs
  Production by $e^+ e^- \to Z H$ in the {Weinberg-Salam} Model}},
  \href{https://doi.org/10.1016/0550-3213(83)90296-1}{\emph{Nucl. Phys. B}
  {\bfseries 216} (1983) 469}.

\bibitem{Kniehl:1991hk}
B.A.~Kniehl, \emph{{Radiative corrections for associated $Z H$ production at
  future $e^{+} e^{-}$ colliders}},
  \href{https://doi.org/10.1007/BF01561297}{\emph{Z. Phys. C} {\bfseries 55}
  (1992) 605}.

\bibitem{Denner:1992bc}
A.~Denner, J.~Kublbeck, R.~Mertig and M.~Bohm, \emph{{Electroweak radiative
  corrections to e+ e- ---\ensuremath{>} H Z}},
  \href{https://doi.org/10.1007/BF01555523}{\emph{Z. Phys. C} {\bfseries 56}
  (1992) 261}.

\bibitem{Kniehl:1995ra}
B.A.~Kniehl, \emph{{Two loop O (alpha-s G(F) $M($Q$) ^{2)}$ heavy quark
  corrections to the interactions between Higgs and intermediate bosons}},
  \href{https://doi.org/10.1103/PhysRevD.53.6477}{\emph{Phys. Rev. D}
  {\bfseries 53} (1996) 6477}
  [\href{https://arxiv.org/abs/hep-ph/9602304}{{\ttfamily hep-ph/9602304}}].

\bibitem{Frink:1996sv}
A.~Frink, B.A.~Kniehl, D.~Kreimer and K.~Riesselmann, \emph{{Heavy Higgs
  lifetime at two loops}},
  \href{https://doi.org/10.1103/PhysRevD.54.4548}{\emph{Phys. Rev. D}
  {\bfseries 54} (1996) 4548}
  [\href{https://arxiv.org/abs/hep-ph/9606310}{{\ttfamily hep-ph/9606310}}].

\bibitem{Djouadi:1997rj}
A.~Djouadi, P.~Gambino and B.A.~Kniehl, \emph{{Two loop electroweak heavy
  fermion corrections to Higgs boson production and decay}},
  \href{https://doi.org/10.1016/S0550-3213(98)00147-3}{\emph{Nucl. Phys. B}
  {\bfseries 523} (1998) 17}
  [\href{https://arxiv.org/abs/hep-ph/9712330}{{\ttfamily hep-ph/9712330}}].

\bibitem{Kniehl:2012rz}
B.A.~Kniehl and O.L.~Veretin, \emph{{Low-mass Higgs decays to four leptons at
  one loop and beyond}},
  \href{https://doi.org/10.1103/PhysRevD.86.053007}{\emph{Phys. Rev. D}
  {\bfseries 86} (2012) 053007}
  [\href{https://arxiv.org/abs/1206.7110}{{\ttfamily 1206.7110}}].

\bibitem{Gong:2016jys}
Y.~Gong, Z.~Li, X.~Xu, L.L.~Yang and X.~Zhao, \emph{{Mixed QCD-EW corrections
  for Higgs boson production at $e^+e^-$ colliders}},
  \href{https://doi.org/10.1103/PhysRevD.95.093003}{\emph{Phys. Rev. D}
  {\bfseries 95} (2017) 093003}
  [\href{https://arxiv.org/abs/1609.03955}{{\ttfamily 1609.03955}}].

\bibitem{Sun:2016bel}
Q.-F.~Sun, F.~Feng, Y.~Jia and W.-L.~Sang, \emph{{Mixed electroweak-QCD
  corrections to e+e-\textrightarrow{}HZ at Higgs factories}},
  \href{https://doi.org/10.1103/PhysRevD.96.051301}{\emph{Phys. Rev. D}
  {\bfseries 96} (2017) 051301}
  [\href{https://arxiv.org/abs/1609.03995}{{\ttfamily 1609.03995}}].

\bibitem{Chen:2018xau}
W.~Chen, F.~Feng, Y.~Jia and W.-L.~Sang, \emph{{Mixed electroweak-QCD
  corrections to $e^+e^-\to \mu^+\mu^- H$ at CEPC with finite-width effect}},
  \href{https://doi.org/10.1088/1674-1137/43/1/013108}{\emph{Chin. Phys. C}
  {\bfseries 43} (2019) 013108}
  [\href{https://arxiv.org/abs/1811.05453}{{\ttfamily 1811.05453}}].

\bibitem{Wang:2019fxh}
Y.~Wang, X.~Xu and L.L.~Yang, \emph{{Two-loop triangle integrals with 4 scales
  for the $HZV$ vertex}},
  \href{https://doi.org/10.1103/PhysRevD.100.071502}{\emph{Phys. Rev. D}
  {\bfseries 100} (2019) 071502}
  [\href{https://arxiv.org/abs/1905.11463}{{\ttfamily 1905.11463}}].

\bibitem{Rizzo:1980gz}
T.G.~Rizzo, \emph{{Decays of Heavy Higgs Bosons}},
  \href{https://doi.org/10.1103/PhysRevD.22.722}{\emph{Phys. Rev. D} {\bfseries
  22} (1980) 722}.

\bibitem{Pocsik:1980ta}
G.~Pocsik and T.~Torma, \emph{{On the Decays of Heavy Higgs Bosons}},
  \href{https://doi.org/10.1007/BF01427913}{\emph{Z. Phys. C} {\bfseries 6}
  (1980) 1}.

\bibitem{Keung:1984hn}
W.-Y.~Keung and W.J.~Marciano, \emph{{HIGGS SCALAR DECAYS: H ---\ensuremath{>}
  W+- X}}, \href{https://doi.org/10.1103/PhysRevD.30.248}{\emph{Phys. Rev. D}
  {\bfseries 30} (1984) 248}.

\bibitem{Frederix:2018nkq}
R.~Frederix, S.~Frixione, V.~Hirschi, D.~Pagani, H.S.~Shao and M.~Zaro,
  \emph{{The automation of next-to-leading order electroweak calculations}},
  \href{https://doi.org/10.1007/JHEP07(2018)185}{\emph{JHEP} {\bfseries 07}
  (2018) 185} [\href{https://arxiv.org/abs/1804.10017}{{\ttfamily
  1804.10017}}].

\bibitem{Smirnov:2019qkx}
A.V.~Smirnov and F.S.~Chuharev, \emph{{FIRE6: Feynman Integral REduction with
  Modular Arithmetic}},
  \href{https://doi.org/10.1016/j.cpc.2019.106877}{\emph{Comput. Phys. Commun.}
  {\bfseries 247Â } (2020) 106877}
  [\href{https://arxiv.org/abs/1901.07808}{{\ttfamily 1901.07808}}].

\bibitem{Lee:2013mka}
R.N.~Lee, \emph{{LiteRed 1.4: a powerful tool for reduction of multiloop
  integrals}}, \href{https://doi.org/10.1088/1742-6596/523/1/012059}{\emph{J.
  Phys. Conf. Ser.} {\bfseries 523} (2014) 012059}
  [\href{https://arxiv.org/abs/1310.1145}{{\ttfamily 1310.1145}}].

\bibitem{Klappert:2020nbg}
J.~Klappert, F.~Lange, P.~Maierh\"ofer and J.~Usovitsch, \emph{{Integral
  reduction with Kira 2.0 and finite field methods}},
  \href{https://doi.org/10.1016/j.cpc.2021.108024}{\emph{Comput. Phys. Commun.}
  {\bfseries 266} (2021) 108024}
  [\href{https://arxiv.org/abs/2008.06494}{{\ttfamily 2008.06494}}].

\bibitem{DiVita:2017xlr}
S.~Di~Vita, P.~Mastrolia, A.~Primo and U.~Schubert, \emph{{Two-loop master
  integrals for the leading QCD corrections to the Higgs coupling to a $W$ pair
  and to the triple gauge couplings $ZWW$ and $\gamma^*WW$}},
  \href{https://doi.org/10.1007/JHEP04(2017)008}{\emph{JHEP} {\bfseries 04}
  (2017) 008} [\href{https://arxiv.org/abs/1702.07331}{{\ttfamily
  1702.07331}}].

\bibitem{Janowski:2021yvz}
T.~Janowski, B.~Pullin and R.~Zwicky, \emph{{Charged and Neutral $\bar
  B_{u,d,s} \to \gamma$ Form Factors from Light Cone Sum Rules at NLO}},
  \href{https://arxiv.org/abs/2106.13616}{{\ttfamily 2106.13616}}.

\bibitem{Kotikov:1990kg}
A.V.~Kotikov, \emph{{Differential equations method: New technique for massive
  Feynman diagrams calculation}},
  \href{https://doi.org/10.1016/0370-2693(91)90413-K}{\emph{Phys. Lett. B}
  {\bfseries 254} (1991) 158}.

\bibitem{Kotikov:1991pm}
A.V.~Kotikov, \emph{{Differential equation method: The Calculation of N point
  Feynman diagrams}},
  \href{https://doi.org/10.1016/0370-2693(91)90536-Y}{\emph{Phys. Lett. B}
  {\bfseries 267} (1991) 123}.

\bibitem{Remiddi:1997ny}
E.~Remiddi, \emph{{Differential equations for Feynman graph amplitudes}},
  {\emph{Nuovo Cim. A} {\bfseries 110} (1997) 1435}
  [\href{https://arxiv.org/abs/hep-th/9711188}{{\ttfamily hep-th/9711188}}].

\bibitem{Gehrmann:1999as}
T.~Gehrmann and E.~Remiddi, \emph{{Differential equations for two loop four
  point functions}},
  \href{https://doi.org/10.1016/S0550-3213(00)00223-6}{\emph{Nucl. Phys. B}
  {\bfseries 580} (2000) 485}
  [\href{https://arxiv.org/abs/hep-ph/9912329}{{\ttfamily hep-ph/9912329}}].

\bibitem{Argeri:2007up}
M.~Argeri and P.~Mastrolia, \emph{{Feynman Diagrams and Differential
  Equations}}, \href{https://doi.org/10.1142/S0217751X07037147}{\emph{Int. J.
  Mod. Phys. A} {\bfseries 22} (2007) 4375}
  [\href{https://arxiv.org/abs/0707.4037}{{\ttfamily 0707.4037}}].

\bibitem{MullerStach:2012mp}
S.~M\"uller-Stach, S.~Weinzierl and R.~Zayadeh, \emph{{Picard-Fuchs equations
  for Feynman integrals}},
  \href{https://doi.org/10.1007/s00220-013-1838-3}{\emph{Commun. Math. Phys.}
  {\bfseries 326} (2014) 237}
  [\href{https://arxiv.org/abs/1212.4389}{{\ttfamily 1212.4389}}].

\bibitem{Henn:2013pwa}
J.M.~Henn, \emph{{Multiloop integrals in dimensional regularization made
  simple}}, \href{https://doi.org/10.1103/PhysRevLett.110.251601}{\emph{Phys.
  Rev. Lett.} {\bfseries 110} (2013) 251601}
  [\href{https://arxiv.org/abs/1304.1806}{{\ttfamily 1304.1806}}].

\bibitem{Henn:2014qga}
J.M.~Henn, \emph{{Lectures on differential equations for Feynman integrals}},
  \href{https://doi.org/10.1088/1751-8113/48/15/153001}{\emph{J. Phys. A}
  {\bfseries 48} (2015) 153001}
  [\href{https://arxiv.org/abs/1412.2296}{{\ttfamily 1412.2296}}].

\bibitem{Ablinger:2015tua}
J.~Ablinger, A.~Behring, J.~Bl\"umlein, A.~De~Freitas, A.~von Manteuffel and
  C.~Schneider, \emph{{Calculating Three Loop Ladder and V-Topologies for
  Massive Operator Matrix Elements by Computer Algebra}},
  \href{https://doi.org/10.1016/j.cpc.2016.01.002}{\emph{Comput. Phys. Commun.}
  {\bfseries 202} (2016) 33}
  [\href{https://arxiv.org/abs/1509.08324}{{\ttfamily 1509.08324}}].

\bibitem{Adams:2017tga}
L.~Adams, E.~Chaubey and S.~Weinzierl, \emph{{Simplifying Differential
  Equations for Multiscale Feynman Integrals beyond Multiple Polylogarithms}},
  \href{https://doi.org/10.1103/PhysRevLett.118.141602}{\emph{Phys. Rev. Lett.}
  {\bfseries 118} (2017) 141602}
  [\href{https://arxiv.org/abs/1702.04279}{{\ttfamily 1702.04279}}].

\bibitem{Bosma:2017hrk}
J.~Bosma, K.J.~Larsen and Y.~Zhang, \emph{{Differential equations for loop
  integrals in Baikov representation}},
  \href{https://doi.org/10.1103/PhysRevD.97.105014}{\emph{Phys. Rev. D}
  {\bfseries 97} (2018) 105014}
  [\href{https://arxiv.org/abs/1712.03760}{{\ttfamily 1712.03760}}].

\bibitem{Chen:2020uyk}
J.~Chen, X.~Jiang, X.~Xu and L.L.~Yang, \emph{{Constructing canonical Feynman
  integrals with intersection theory}},
  \href{https://doi.org/10.1016/j.physletb.2021.136085}{\emph{Phys. Lett. B}
  {\bfseries 814} (2021) 136085}
  [\href{https://arxiv.org/abs/2008.03045}{{\ttfamily 2008.03045}}].

\bibitem{Baikov:1996iu}
P.A.~Baikov, \emph{{Explicit solutions of the multiloop integral recurrence
  relations and its application}},
  \href{https://doi.org/10.1016/S0168-9002(97)00126-5}{\emph{Nucl. Instrum.
  Meth. A} {\bfseries 389} (1997) 347}
  [\href{https://arxiv.org/abs/hep-ph/9611449}{{\ttfamily hep-ph/9611449}}].

\bibitem{Mizera:2017rqa}
S.~Mizera, \emph{{Scattering Amplitudes from Intersection Theory}},
  \href{https://doi.org/10.1103/PhysRevLett.120.141602}{\emph{Phys. Rev. Lett.}
  {\bfseries 120} (2018) 141602}
  [\href{https://arxiv.org/abs/1711.00469}{{\ttfamily 1711.00469}}].

\bibitem{Mastrolia:2018uzb}
P.~Mastrolia and S.~Mizera, \emph{{Feynman Integrals and Intersection Theory}},
  \href{https://doi.org/10.1007/JHEP02(2019)139}{\emph{JHEP} {\bfseries 02}
  (2019) 139} [\href{https://arxiv.org/abs/1810.03818}{{\ttfamily
  1810.03818}}].

\bibitem{Frellesvig:2019kgj}
H.~Frellesvig, F.~Gasparotto, S.~Laporta, M.K.~Mandal, P.~Mastrolia,
  L.~Mattiazzi et~al., \emph{{Decomposition of Feynman Integrals on the Maximal
  Cut by Intersection Numbers}},
  \href{https://doi.org/10.1007/JHEP05(2019)153}{\emph{JHEP} {\bfseries 05}
  (2019) 153} [\href{https://arxiv.org/abs/1901.11510}{{\ttfamily
  1901.11510}}].

\bibitem{Frellesvig:2019uqt}
H.~Frellesvig, F.~Gasparotto, M.K.~Mandal, P.~Mastrolia, L.~Mattiazzi and
  S.~Mizera, \emph{{Vector Space of Feynman Integrals and Multivariate
  Intersection Numbers}},
  \href{https://doi.org/10.1103/PhysRevLett.123.201602}{\emph{Phys. Rev. Lett.}
  {\bfseries 123} (2019) 201602}
  [\href{https://arxiv.org/abs/1907.02000}{{\ttfamily 1907.02000}}].

\bibitem{Mizera:2019vvs}
S.~Mizera and A.~Pokraka, \emph{{From Infinity to Four Dimensions: Higher
  Residue Pairings and Feynman Integrals}},
  \href{https://doi.org/10.1007/JHEP02(2020)159}{\emph{JHEP} {\bfseries 02}
  (2020) 159} [\href{https://arxiv.org/abs/1910.11852}{{\ttfamily
  1910.11852}}].

\bibitem{Weinzierl:2020xyy}
S.~Weinzierl, \emph{{On the computation of intersection numbers for twisted
  cocycles}},  \href{https://arxiv.org/abs/2002.01930}{{\ttfamily 2002.01930}}.

\bibitem{Frellesvig:2020qot}
H.~Frellesvig, F.~Gasparotto, S.~Laporta, M.K.~Mandal, P.~Mastrolia,
  L.~Mattiazzi et~al., \emph{{Decomposition of Feynman Integrals by
  Multivariate Intersection Numbers}},
  \href{https://doi.org/10.1007/JHEP03(2021)027}{\emph{JHEP} {\bfseries 03}
  (2021) 027} [\href{https://arxiv.org/abs/2008.04823}{{\ttfamily
  2008.04823}}].

\bibitem{Goncharov:1998kja}
A.B.~Goncharov, \emph{{Multiple polylogarithms, cyclotomy and modular
  complexes}}, \href{https://doi.org/10.4310/MRL.1998.v5.n4.a7}{\emph{Math.
  Res. Lett.} {\bfseries 5} (1998) 497}
  [\href{https://arxiv.org/abs/1105.2076}{{\ttfamily 1105.2076}}].

\bibitem{Besier:2018jen}
M.~Besier, D.~Van~Straten and S.~Weinzierl, \emph{{Rationalizing roots: an
  algorithmic approach}},
  \href{https://doi.org/10.4310/CNTP.2019.v13.n2.a1}{\emph{Commun. Num. Theor.
  Phys.} {\bfseries 13} (2019) 253}
  [\href{https://arxiv.org/abs/1809.10983}{{\ttfamily 1809.10983}}].

\bibitem{Besier:2019kco}
M.~Besier, P.~Wasser and S.~Weinzierl, \emph{{RationalizeRoots: Software
  Package for the Rationalization of Square Roots}},
  \href{https://doi.org/10.1016/j.cpc.2020.107197}{\emph{Comput. Phys. Commun.}
  {\bfseries 253} (2020) 107197}
  [\href{https://arxiv.org/abs/1910.13251}{{\ttfamily 1910.13251}}].

\bibitem{Bauer:2000cp}
C.W.~Bauer, A.~Frink and R.~Kreckel, \emph{{Introduction to the GiNaC framework
  for symbolic computation within the C++ programming language}},
  \href{https://doi.org/10.1006/jsco.2001.0494}{\emph{J. Symb. Comput.}
  {\bfseries 33} (2002) 1} [\href{https://arxiv.org/abs/cs/0004015}{{\ttfamily
  cs/0004015}}].

\bibitem{Vollinga:2005pk}
J.~Vollinga, \emph{{GiNaC: Symbolic computation with C++}},
  \href{https://doi.org/10.1016/j.nima.2005.11.155}{\emph{Nucl. Instrum. Meth.
  A} {\bfseries 559} (2006) 282}
  [\href{https://arxiv.org/abs/hep-ph/0510057}{{\ttfamily hep-ph/0510057}}].

\bibitem{Vollinga:2004sn}
J.~Vollinga and S.~Weinzierl, \emph{{Numerical evaluation of multiple
  polylogarithms}},
  \href{https://doi.org/10.1016/j.cpc.2004.12.009}{\emph{Comput. Phys. Commun.}
  {\bfseries 167} (2005) 177}
  [\href{https://arxiv.org/abs/hep-ph/0410259}{{\ttfamily hep-ph/0410259}}].

\bibitem{Naterop:2019xaf}
L.~Naterop, A.~Signer and Y.~Ulrich, \emph{{handyG \textemdash{}Rapid numerical
  evaluation of generalised polylogarithms in Fortran}},
  \href{https://doi.org/10.1016/j.cpc.2020.107165}{\emph{Comput. Phys. Commun.}
  {\bfseries 253} (2020) 107165}
  [\href{https://arxiv.org/abs/1909.01656}{{\ttfamily 1909.01656}}].

\bibitem{Borowka:2017idc}
S.~Borowka, G.~Heinrich, S.~Jahn, S.P.~Jones, M.~Kerner, J.~Schlenk et~al.,
  \emph{{pySecDec: a toolbox for the numerical evaluation of multi-scale
  integrals}}, \href{https://doi.org/10.1016/j.cpc.2017.09.015}{\emph{Comput.
  Phys. Commun.} {\bfseries 222} (2018) 313}
  [\href{https://arxiv.org/abs/1703.09692}{{\ttfamily 1703.09692}}].

\bibitem{Hahn:2000kx}
T.~Hahn, \emph{{Generating Feynman diagrams and amplitudes with FeynArts 3}},
  \href{https://doi.org/10.1016/S0010-4655(01)00290-9}{\emph{Comput. Phys.
  Commun.} {\bfseries 140} (2001) 418}
  [\href{https://arxiv.org/abs/hep-ph/0012260}{{\ttfamily hep-ph/0012260}}].

\bibitem{Djouadi:1993ss}
A.~Djouadi and P.~Gambino, \emph{{Electroweak gauge bosons selfenergies:
  Complete QCD corrections}},
  \href{https://doi.org/10.1103/PhysRevD.49.3499}{\emph{Phys. Rev. D}
  {\bfseries 49} (1994) 3499}
  [\href{https://arxiv.org/abs/hep-ph/9309298}{{\ttfamily hep-ph/9309298}}].

\bibitem{Dittmaier:2015rxo}
S.~Dittmaier, A.~Huss and C.~Schwinn, \emph{{Dominant mixed QCD-electroweak
  O($\alpha_s\alpha$) corrections to Drell\textendash{}Yan processes in the
  resonance region}},
  \href{https://doi.org/10.1016/j.nuclphysb.2016.01.006}{\emph{Nucl. Phys. B}
  {\bfseries 904} (2016) 216}
  [\href{https://arxiv.org/abs/1511.08016}{{\ttfamily 1511.08016}}].

\bibitem{Zyla:2020zbs}
{\scshape Particle Data Group} collaboration, \emph{{Review of Particle
  Physics}}, \href{https://doi.org/10.1093/ptep/ptaa104}{\emph{PTEP} {\bfseries
  2020} (2020) 083C01}.

\end{thebibliography}\endgroup

\end{document}